\documentclass[final,longbibliography,12pt]{revtex4-1}

\usepackage{amsmath}  
\usepackage{amsfonts} 
\usepackage{graphicx} 
\usepackage{slashed} 

\newcommand{\be}{\begin{equation}}
\newcommand{\ee}{\end{equation}}
\newcommand{\wf}{wavefunction\;}
\newcommand{\wfs}{wavefunctions\;}

\newcommand{\infls}{influence functions\;}
\newcommand{\rmi}{\mathrm{i}}

\newcommand{\RF}{${\mathbb R}^4$}
\newcommand{\LC}{{\mathcal L}}

\newcommand{\fy}{\slashed}

\newcommand{\vt}{\vartheta}

\begin{document}

\title{Feynman-Stueckelberg electroweak interactions and isospin entanglement}

\author{A. F. Bennett}
\email{bennetan@oregonstate.edu} 
\affiliation{College of Earth, Ocean and Atmospheric Sciences\\
Oregon State University\\104 CEOAS Administration Building\\ Corvallis, OR 97331-5503, USA} 

\date{\today}

\begin{abstract}

Entanglement in  Quantum Field Theory is restricted to spacelike separations to the order of the Compton wavelength $\hbar/mc$ (e.g., S. J. Summers and  R. Werner, {\it J. Math. Phys.}, {\bf 28}, 10,2440-2447, (1987)). Yet spin entanglement of electrons across macroscopic distances has been observed by Hensen { \it et al.} ({\it Nature}, {\bf 526}, doi:10.1038/nature/15759, (2015)).  The parametrized relativistic quantum mechanics of Feynman and Stueckelberg  admits spin singlets, across arbitrary separations, by providing a single covariant wave equation for tensor products of two Dirac spinors (A. F. Bennett, {\it Ann. Phys.} {\bf 345}, 1-16 (2014)). The formalism is extended here from  quantum electrodynamics to the electroweak interaction. A relativistic Bell's inequality for Dirac spinors  is extended here to weak isospin.

\end{abstract}
 \maketitle

\section{Introduction}\label{S:int}
The parametrized relativistic quantum mechanics (hereafter PM) of R. Feynman \cite{Alv98} and C. Stueckelberg \cite{Stu_Works} can with one exception represent every phenomenon of Quantum Electrodynamics  \cite{Benn2014,Benn2015,Benn2016,BennPairs15arX}. The sole exception is  anti-bunching in quantum optics at very low intensity \cite{Thorn04,Fox06}. On the other hand PM, unlike Quantum Field Theory (QFT), admits electron spin entanglement across macroscopic proper distances as has now been observed \cite{hensen2015}. PM is extended here to the electroweak interaction. The following sections include a brief statement of the Standard Model (SM) freely referring to a standard modern monograph \cite{Dono14} both for detail and for notation. The PM representation of the SM is identical to QFT, except that (i) the dependent variables are c-valued wavefunctions rather than fields of operators on Fock space, and (ii) the Feynman-Stueckelberg parameter is introduced into the Higgs-fermion couplings.  The parameter $\tau$ has physical reality since it explains quantum interference in local coordinate time \cite{Ho06}. Finally, the Pauli-Lubanski matrices are used to extend the relativistic Bell's inequality  for Dirac spinors \cite{BennPairs15arX} to weak isospin.

\section{The Electroweak Lagrangian}\label{S:EWLA}

\subsection{the Weinberg-Salam Lagrange density}
\label{S:WSLd}

The representation here of the Standard Model  is in the form of parametrized relativistic quantum mechanics. The   \wfs for the fermions  all depend upon the same real parameter $\tau$\ having the range $-\infty<\tau<\infty$\,. For example, a Dirac 4--spinor \wf becomes $\psi(x,\tau)$, where $x_\mu$ (for $\mu=0,1,2,3$)  or simply $x$ is an event in spacetime. The Lorentz metric $g^{\mu\nu}=\mathrm {diag}(-1,1,1,1)$ is restricted to \RF \,, that is, it does not  include $\tau$\,.

The gauge bosons $\mathbf{W}_\mu(x), B_\mu(x)$ and the Higgs field $\Phi(x)$ are classical fields, all of which are independent of $\tau$\,. Their Lagrange densities  $\LC_G(x)$ and $\LC_{HG}(x)$ are identical to those of the SM. See   \cite[p60]{Dono14}. The notation therein is conventional and is closely followed here. For example, the Higgs self-interaction in $\mathcal{L}_{HG}$ is
\be\label{Hpot}
V(\Phi)=-\mu^2\Phi^\dag\Phi +\lambda(\Phi^\dag\Phi)^2\,.
\ee

 The fermion wavefunctions  $\Psi(x,\tau)$ are, as indicated, dependent upon $\tau$.  Their Lagrange densities $\LC_F(x,\tau)$ are identical  to  those of  the SM, see again \cite[p60]{Dono14}. The fermion Lagrangians  include averaging over $\tau$, as well as summing over spacetime exactly as in the SM. That is, 
\begin{multline}\label{Lags}
\LC_{WS}(x)=\LC_G(x)+\LC_{HG} (x) \\ +\lim_{T \to \infty}\frac{1}{T}\int^{T/2}_{-T/2} \biggl(\LC_F(x,\tau)+\LC_{HF}(x,\tau)\biggr)d\tau\,.
\end{multline}
It follows immediately from (\ref{Lags}) that the gauge fields and the Higgs boson are supported by $\tau$-averaged fermion currents. 

As will be seen in (\ref{Delta}) below, the only departure from the SM  is in the Lagrange density $\LC_{HF}$ for the Higgs-fermion couplings. The  Higgs ground state operator $-\rmi \partial _\tau$ replaces the Higgs ground state c-number   $v/\sqrt{2}=\mu/\sqrt{2\lambda}$, in SM  notation \cite[p63]{Dono14} \footnote{ $v=M_1/ \sqrt{h} $ in the notation of \cite{Wei67}}. The c-number is not so replaced elsewhere.

\subsection{gauge bosons}\label{S:Gauge}
 The gauge bosons  $\mathbf{W}_\mu$ and $B_\mu$\,, being independent of $\tau$\,, are Standard. Their contributions to the mass Lagrangian $\mathcal{L}_{mass}$ are Standard, as in \cite[p62]{Dono14}. In particular, the bare mass $M_W$ of the charged vector bosons $W\mu^\pm$ has the Standard value $M_W=g_2\mu/2\sqrt{\lambda}$ where $g_2$ is the $SU(2)_L$  coupling constant. The weak mixing angle $\theta_W$ is Standard ($\tan\theta_W=g_1/g_2$, where $g_1$ is the $U(1)$ coupling constant), as are the masses for  the photon $A_\mu$ ($M_\gamma=0$) and the massive neutral boson $Z_\mu $ ($M_Z=M_W/ \theta_W) $ \cite[p62-63]{Dono14}.

 \subsection{fermions} \label{S:Hferm}
The Lagrange density for Higgs couplings to the first generation of quarks and leptons is  $\LC_{HF}=\LC_{Hq}+\LC_{Hl}$\,. For up and down quarks,
\be\label{Hq}
\LC_{Hq}=-g_u\overline{q}_L\widetilde{\Delta}u_R-g_d\overline{q}_L\Delta d_R+\mathrm {h.c.}\,,
\ee
where
\be\label{Delta}
\Delta=\left(
\begin{matrix}
0\\
-\rmi \partial_\tau\end{matrix}\right)+
\left(
\begin{matrix}
\varphi^+\\
\phi^0\end{matrix}\right)\,,
\ee
where $\varphi^+$ and $\varphi^0$ are spin-zero Higgs wavefunctions ``with electric charge assignments as indicated". \cite{Dono14}, and where $\widetilde{\Delta}=\rmi\tau_2\Delta^*$\,. Here, unfortunately,  $\tau_2$ represents one of the three isospin matrices. Again, the Higgs vacuum operator $-\rmi \partial _\tau$ in (\ref{Delta})  replaces the Higgs vacuum  c-number $\vt \equiv v/\sqrt{2}$ of the SM. The coupling constants $g_u$ and $g_d$ are related to the rest masses of particles by $m_u=g_u\vt$ and $m_d=g_d\vt$ respectively.  Combining $\LC_F$ and $\LC_{Hq}$ establishes that  a free  up-quark $u$, for example, obeys the parametrized Dirac wave equation
\be\label{pDir}
\bigl(\fy{\partial}+ g_u \partial_\tau \bigr)u=0\,.
\ee
For a plane wave  $u(x,\tau)=u(0,0)\exp[\rmi(p\cdot x + \varpi_p \tau)]$, the dispersion relation is $p \cdot p =-g_u^2\varpi^2_p$. It is now assumed that  a fermion  is  ``on shell" if  $\varpi_p =\vt=v/\sqrt{2}$\,.  Hence a free on-shell up-quark  $u(x,\tau)$ satisfies  the Dirac equation
\be\label{Dir}
(\rmi \fy{\partial}-m_u)u=0\,,
\ee
where $m_u=g_u \theta$ is the up-quark mass. Details of the free wavefunctions, discrete symmetries and \infls may be found in \cite{Benn2014}. The normalization factor here for the spinor amplitudes of a free electron \wf is $\sqrt{(E_p+m_p)/2g_e m_p}$ where $E_p=p^0$ and $m_p=g_e\varpi_p$\,. A free particle on mass shell propagates on mass shell.  If an initial particle in a scattering process is on mass shell $(\varpi_p=\vt)$ then, as a consequence of the scattering field being independent of $\tau$, the final particle is also on mass shell. 

 For any same-generation fermion doublets $\Psi(x,\tau)$ and $\Upsilon(x,\tau)$  satisfying 
 \be\label{doublets}
 \Big(\fy{D}+{\mathbf g}\,\partial_\tau\Big)\Psi(x,\tau)= \Big(\fy{D}+{\mathbf g}\,\partial_\tau\Big)\Upsilon(x,\tau)=0\,,
 \ee
it may be shown that 
\be\label{consf}
\frac{\partial }{\partial x^{\mu}}(\overline{\Upsilon}\gamma^{\mu}\Psi)+\frac{\partial }{\partial \tau} \overline{\Upsilon}\,{\mathbf g}\, \Psi=0\,.
\ee
In the case of the first quark generation, for example, ${\mathbf g}={\mathbf g}^{(1)}$ is the diagonal matrix $ \mathrm {diag} \,(g_u,g_d)$\,.  It follows from (\ref{consf}) that the invariant bilinear form $ \int \overline{\Upsilon}\,{\mathbf g}\, \Psi d^{\,4}x$ is independent of $\tau$\,. The parametrized wave equation for two fermion doublets is
\be\label{2Dirac}
 \Big(\fy{D }_x\otimes {\mathbf 1}+{\mathbf 1}\otimes \fy{D}_y\Big)\Theta(x,y,\tau)+\partial_\tau \Big({\mathbf g}^{(1)}\otimes{\mathbf g}^{(2)}\Theta(x,y,\tau) \Big)=0\,,
\ee
where the \wf $\Theta(x,y,\tau)$ is in general a sum of tensor products of  doublets such as  the \wfs $\Psi(x,\tau)$ and $\Upsilon(y,\tau)$. There are two coordinate times in (\ref{2Dirac}), namely $x^0$ and $y^0$. If the two-particle \wf $\Theta$ and hence (\ref{2Dirac}) had not been parametrized, then  integration with respect to either $x^0$ or $y^0$ would violate covariance. The parametrization of $\Theta$\,, and  integration of (\ref{2Dirac}) with respect to the parameter $\tau$ preserves covariance.


\section{weak isospin entanglement}\label{S:weak}
The relativistic Bell's inequality for a singlet of Dirac spinors is an elementary  paraphrase   of the non-relativistic development for a singlet of Pauli spinors \cite[\S12.2]{WeiQM13}. The covariant spin operator for Dirac spinors is $\gamma_5\fy{a}$ where $a^\mu a_\mu=-1$, replacing the  Pauli spin operator $\sigma_j a_j$ where $a_j a_j=1$. The relativistic Bell's inequality for weak isospin requires only the definition of a covariant operator for weak isospin. The definition is as follows.

The isospin doublets $\Psi(x,\tau)$ and $\Upsilon(x,\tau)$  for same-generation fermions must respect their spin-statistics, that is, a two-doublet \wf $\Theta$ must be of the form
\be\label{fd}
\Theta(x,y,\tau) = \frac{1}{\sqrt{2}}\Big(\Psi(x,\tau) \otimes \Upsilon(y,\tau) - \Upsilon(x,\tau)\otimes\Psi(y,\tau)\Big)\,.
\ee

The two-doublet state is manifestly entangled in spacetime, and the parametrized wave equation (\ref{2Dirac})
in no way restricts the separation $x-y$ of this entanglement.  The $SU(2)$  spinor basis is $\tau^\mu=\{\mathbf{1}, \tau_1, \tau_2, \tau_3\}$ where again the $\tau_j$ for  $ j=1,2,3$\, are the Pauli isospin matrices.  The dual basis is $\hat{\tau}^\mu=\{\mathbf{1}, -\tau_1,- \tau_2, -\tau_3\}$\,. An invariant inner product  is provided, again, by $\int \overline{\Psi}\mathbf{g}\Upsilon d^{\,4}x$ for all positive-energy isospin doublets $\Psi(x,\tau)$ and $\Upsilon(x,\tau)$\,. The two Dirac spinor components of the doublets transform independently and covariantly in the usual way \cite[Ch2]{BjDr64}, while the doublets transform in the $SU(2)$ representation. 
 Next, the Dirac matrices $\gamma^\mu$ are replaced with the  Pauli-Lubanski matrices $X^\mu(p)= \rmi L^{\mu \nu}p_\nu$\,, where the objects $L^{\mu\nu}=(\rmi/4)(\tau^\mu\hat{\tau}^\nu-\tau^\nu\hat{\tau}^\mu)$ generate the $SU(2)$  representation of the Lorentz group \cite{IZ}. The isospin operator is then $-(2/m_p)a_\mu X^\mu(p)$, for any $a$ such that  $p\cdot a=0$ and $a\cdot a=-1$\,. 
 
 The spatial extent of a static vector boson is its Compton wavelength and so the feasibility of a ``weak Stern-Gerlach apparatus" seems remote, but parametrized relativistic quantum mechanics does admit unrestricted entanglement of left-handed fermion doublets.

\section*{References}
\bibliography{paraDirac_Bib}

\end{document}